# Robust Image Classification: Defensive Strategies against FGSM and PGD Adversarial Attacks

Hetvi Waghela
Department of Data Science
Praxis Business School
Kolkata, INDIA
email: waghelah@acm.org

Jaydip Sen
Department of Data Science
Praxis Business School
Kolkata, INDIA
email: jaydip.sen@acm.org

Sneha Rakshit
Department of Data Science
Praxis Business School
Kolkata, INDIA
email: srakshit149@gmail.com

*Abstract*—Adversarial attacks, particularly the Fast Gradient Sign Method (FGSM) and Projected Gradient Descent (PGD) pose significant threats to the robustness of deep learning models in image classification. This paper explores and refines defense mechanisms against these attacks to enhance the resilience of neural networks. We employ a combination of adversarial training and innovative preprocessing techniques, aiming to mitigate the impact of adversarial perturbations. Our methodology involves modifying input data before classification and investigating different model architectures and training strategies. Through rigorous evaluation of benchmark datasets, we demonstrate the effectiveness of our approach in defending against FGSM and PGD attacks. Our results show substantial improvements in model robustness compared to baseline methods, highlighting the potential of our defense strategies in real-world applications. This study contributes to the ongoing efforts to develop secure and reliable machine learning systems, offering practical insights and paving the way for future research in adversarial defense. By bridging theoretical advancements and practical implementation, we aim to enhance the trustworthiness of AI applications in safety-critical domains.

## I. Introduction

Over the past few years, deep learning has emerged as a revolutionary technology, propelling notable progress in diverse domains like computer vision, natural language processing, and autonomous systems. Image classification is a prominent application of deep learning, with convolutional neural networks (CNNs) showcasing exceptional performance, frequently exceeding human-level accuracy on complex datasets. These achievements are underpinned by sophisticated architectures and large-scale datasets that enable models to learn complex patterns and representations. However, despite their impressive performance, deep learning models are not without vulnerabilities, particularly to adversarial attacks that can compromise their reliability and security.

Adversarial attacks entail purposeful alterations to provide input data that misleads deep learning models, resulting in inaccurate predictions. These manipulations, known as adversarial perturbations, are typically small and carefully crafted to be imperceptible to the human eye. This subtlety makes adversarial attacks particularly dangerous, as the altered inputs appear normal to humans but can cause models to fail catastrophically. Among the various adversarial attack techniques, FGSM) and PGD attacks are two of the most popular widely studied methods.

FGSM, introduced by Goodfellow et al. in 2014, is a single-step attack that generates adversarial examples by adding a perturbation along the gradient direction of the loss function concerning the input [1]. The perturbation is scaled by a factor to ensure it remains small but impactful. FGSM's efficiency and straightforward implementation have made it a widely favored option for adversarial attacks, as it can quickly generate examples that significantly degrade model performance. PGD, on the other hand, is a multi-step variant of FGSM [2]. It iteratively applies the FGSM attack within a specified perturbation bound, making it a more potent and iterative approach to generating adversarial examples. PGD is regarded as among the most potent first-order adversarial attacks due to its iterative nature, which allows it to find more effective perturbations.

Deep learning models' vulnerability to these attacks has serious consequences, particularly in high-stakes areas such as self-driving vehicles, medical diagnostics, and digital security. For instance, an adversarial attack on an autonomous vehicle's vision system could lead to misinterpretation of road signs, resulting in dangerous driving behavior. Similarly, in medical imaging, adversarial examples could cause diagnostic models to misclassify benign lesions as malignant, or vice versa, leading to incorrect treatment decisions. The potential for such high-stakes failures underscores the urgency of developing robust defense mechanisms against adversarial attacks.

Research into defending against adversarial attacks has grown rapidly, with various strategies being proposed to enhance the robustness of neural networks [3][4][5][6][7][8][9][10]. Defense mechanisms can be broadly categorized into several approaches, including adversarial training, input preprocessing, model architecture modifications, and detection mechanisms.

Adversarial training, which involves augmenting the training dataset with adversarial examples, is one of the most straightforward and widely used defense strategies. By training the model on both clean and adversarially perturbed inputs, it learns to recognize and resist such perturbations. However, adversarial training is computationally expensive, requiring significant additional resources to generate and incorporate adversarial examples during training. Moreover, it may not generalize well to new types of attacks not seen during training.

Input preprocessing techniques consist of altering the data in a manner that removes or mitigates adversarial perturbations before it is fed into the classifier. Methods such as input denoising, image compression, and adversarial example detection fall under this category. These techniques can be effective but often introduce trade-offs in terms of computational overhead and potential degradation of the input data's quality.

Model architecture modifications involve designing neural network structures that are inherently more robust to adversarial attacks. This can include using specialized activation functions, implementing robust training algorithms, or incorporating additional layers specifically designed to detect and neutralize adversarial perturbations. While promising, these approaches often require significant changes to existing models and may not be compatible with all architectures.

Detection mechanisms aim to identify adversarial examples before they can affect the model's predictions. These methods typically involve training a secondary model or using statistical techniques to detect discrepancies between normal and adversarial inputs. While effective in some cases, detection mechanisms can also suffer from high false positive rates and may struggle to keep up with the evolving nature of adversarial attacks.

The current work aims to address these challenges by exploring and refining defense mechanisms against PGD and FGSM attacks. We focus on enhancing the robustness of image classifiers through a combination of adversarial training and innovative preprocessing techniques. We modify the data prior to presenting it to the classifier as part of our approach, aiming to reduce the impact of adversarial perturbations. Furthermore, we explore the efficacy of employing various model architectures and training strategies to enhance the robustness of neural networks against adversarial attacks.

This paper's structure is as follows. Section II, reviews the relevant adversarial attacks and defense mechanisms, highlighting the strengths and limitations of existing approaches. Section III details the methodology, including a comprehensive explanation of FGSM and PGD attacks and the proposed defense strategies. Section IV, presents the experimental configuration and outcomes, demonstrating the effectiveness of our approach through rigorous evaluation of benchmark datasets. Section V discusses the findings, insights into the effectiveness of our defense mechanisms, and potential areas for future research. Finally, Section VI concludes the paper by including a recapitulation of our contributions and the implications of our proposition for the broader field of machine learning security

## II. RELATED WORK

Within the deep learning community, there has been a substantial interest in both adversarial attacks and defenses, largely due to their implications for the security and robustness of machine learning models. This section reviews key works in this domain, focusing on the foundational theories of adversarial attacks, prominent defense mechanisms, and their respective strengths and limitations.

*Adversarial Attacks:* First, we discuss some important works on adversarial attacks.

The concepts of adversarial examples and the FGSM attack were introduced in [1]. A major strength of this work is its clear demonstration that small, intentional perturbations to input data can lead deep learning models to confidently make inaccurate predictions. However, the scope of the proposed defense is limited as adversarial training can be computationally expensive and may not generalize well to unseen types of attacks.

PGD as a robust method for generating adversarial examples and using it for adversarial training to defend against attacks was proposed in [2]. A key strength of this work is its rigorous theoretical foundation, which demonstrates that adversarial training with PGD significantly enhances the robustness of deep learning models. However, while the defense is effective against first-order attacks like PGD, its performance against more sophisticated or adaptive attacks remains less explored.

A comprehensive analysis of adversarial attacks on image classifiers alongside explorations of defense mechanisms for them have been presented in [4][5]. One strength of the work is its thorough evaluation of different types of attacks, providing valuable insights into their impacts on model performance. However, the proposed defenses may be computationally intensive and might not scale well with larger datasets or more complex models.

A work on critical examination of the efficacy of the existing detection mechanism for adversarial examples is presented in [6]. The authors introduce sophisticated attacks that successfully bypass ten state-of-the-art detection methods, highlighting a significant weakness in current defense strategies. However, the work does not propose any improved detection mechanisms for these attacks.

A novel adversarial attack targeting image captioning models using attention-based optimization techniques has been proposed by some researchers [7]. The effectiveness of this method is rooted in its capability to produce highly effective adversarial examples that exploit the attention mechanism, leading to more precise and impactful perturbations. However, the attack's complexity and computational demands may pose challenges.

Bortsova et al. examine the susceptibility of medical image analysis systems to adversarial attacks, focusing on the impact of these attacks on diagnostic accuracy [8]. The authors explore various factors contributing to vulnerability, such as the choice of model architecture, the diversity of the training dataset, and the presence of domain-specific features. They demonstrate that even small perturbations in input images can cause significant misclassifications, highlighting the inadequacy of current defense mechanisms. Additionally, the authors propose a framework for systematically assessing the robustness of medical AI systems.

Li et al. propose a new adversarial attack method called the Block Gray Adversarial Attack (BGAA), designed to fool image classification neural networks [9]. The BGAA technique involves dividing an image into blocks and applying perturbations in a specific color range, making the attack less noticeable to humans. This method is particularly effective against deep learning models, as it exploits their sensitivity to subtle changes in input data. The authors demonstrate that BGAA can achieve a high attack success rate with minimal perturbations, outperforming traditional adversarial attack methods. The study also emphasizes the need for developing more robust defenses to mitigate such attacks.

*Defense Mechanisms:* The following are some of the well-known defense mechanisms for adversarial attacks proposed in the literature.

The concept of defensive distillation, introduced in [3], aims to bolster the resilience of neural networks against adversarial manipulation. The primary strength of this work is its novel approach that leverages the process of distillation to make the model's predictions less sensitive to small input perturbations, thereby improving resistance to adversarial examples. However, defensive distillation may not be effective against more sophisticated attacks that have emerged since its proposal.

Investigations of the vulnerabilities of large multimodal models to adversarial attacks have been made in [10]. The key strength of this work is its comprehensive analysis of how multimodal models, which integrate visual and textual information, respond to adversarial perturbations, providing valuable insights into their robustness. However, a notable weakness is that the study primarily focuses on specific types of attacks and models, potentially limiting the generalizability of the findings to other multimodal systems of attack strategies.

An ensemble training as a means to augment the resilience of deep neural networks to defend against adversarial samples by incorporating adversarial samples generated from multiple models during training is proposed in [11]. However, the technique is computationally intensive, which could limit its scalability and practical applications in real-world scenarios.

A new defense mechanism inspired by biological systems is introduced to combat adversarial attacks in image classification [12]. The innovative approach, which draws inspiration from biological systems to enhance the robustness of neural networks, potentially offers new avenues for defense. However, the effectiveness of the proposition may vary across different types of attacks and datasets, and the approach might require further validation and optimization.

The work in [13] presents a critical analysis of the effectiveness of various defense mechanisms that rely on gradient obfuscation. However, a weakness of this work is that while it effectively critiques current defenses, it does not propose new robust defense strategies, leaving the challenge of developing truly secure models unresolved.

Chen et al. explore the vulnerability of image classification models to adversarial attacks [14]. The authors highlight various attack methods, such as gradient-based and optimization-based techniques, that can deceive models into making incorrect classifications. They discuss the effectiveness of these attacks across different models and datasets, emphasizing the need for robust defenses. They also review existing defense mechanisms, including adversarial training, defensive distillation, and input transformation techniques, assessing their strengths and limitations.

Li & Cao propose a novel defense strategy against adversarial attacks on image classification models [15]. The authors introduce a sparse denoiser that leverages image label information and pixel guidance to mitigate the effects of adversarial perturbations. The denoiser aims to selectively remove adversarial noise while preserving the essential features necessary for accurate classification. Experimental results demonstrate that this approach effectively improves model robustness against various adversarial attacks.

Entezari & Papalexakis introduce a tensor decomposition-based defense mechanism called Tensorshield [16]. This approach leverages tensor decomposition to separate the essential data structure of images from adversarial perturbations. By focusing on the core components of the tensor representation, Tensorshield aims to reconstruct clean images and mitigate the impact of adversarial noise. The method demonstrates robustness against various adversarial attacks, outperforming traditional defense techniques in preserving classification accuracy. The authors emphasize the potential of tensor-based methods in enhancing the security of image classification systems.

***Input Processing and Model Modification:*** The following are some important works in this area.

A novel preprocessing approach called *feature squeezing* has been developed to identify and alleviate adversarial examples by limiting the input space accessible to potential attackers [17]. However, feature squeezing might not effectively counter all forms of attacks.

A novel approach utilizing Generative Adversarial Networks (GANs) to enhance classifier robustness is proposed in [18]. Through the integration of adversarial examples generated by GANs in the training, the classifier becomes capable of recognizing and classifying such perturbations, improving its resilience. However, while the method shows promise in mitigating adversarial vulnerabilities, further exploration is needed to evaluate its efficacy across diverse datasets and attack strategies.

In [19], a novel approach is proposed leveraging neural fingerprints to identify adversarial examples by capturing distributional discrepancies between clean and perturbed inputs. The method aims to enhance model robustness by preemptively flagging potentially malicious inputs before they are processed by the classifier. However, the efficacy of the scheme could fluctuate based on the intricacy and range of the input data.

A method called randomized smoothing to achieve certified robustness against adversarial attacks is proposed in [20]. In this method, the input data is perturbed with random noise, and the model's predictions are aggregated to provide probabilistic guarantees of robustness. However, the effectiveness of the scheme may diminish with higher-dimensional data.

Huang et al. present a novel defense strategy against adversarial attacks on image classification models [21]. The authors propose using a density-based representation of images to differentiate between genuine data and adversarially perturbed inputs. This representation involves analyzing the distribution of pixel intensities, which helps to identify and filter out anomalies introduced by attacks. The approach enhances model robustness by focusing on the intrinsic characteristics of images rather than the perturbed surface features.

Ren et al. provide an overview of the susceptibilities of deep neural networks to attacks, where slight perturbations to input data can cause significant model misbehavior [22]. It categorizes various types of adversarial attacks, including white-box and black-box attacks, and explores different methods used to create adversarial examples. The authors also discuss defense mechanisms, such as adversarial training, defensive distillation, and input transformation techniques, aimed at improving model robustness. The work also highlights the challenges and limitations of current defense strategies, emphasizing the need for more robust and generalizable solutions. The author also proposes outlines of the designs of more robust deep learning-based image classifiers.

Zoran et al. explore the use of sequential attention mechanisms to improve the robustness of image classification models [23]. The authors propose a novel approach where a model sequentially attends to different parts of an image, refining its understanding and classification decision over time. This method helps the model focus on important image features, reducing susceptibility to adversarial attacks and improving generalization. The authors also demonstrate that sequential attention models achieve better performance and robustness compared to traditional convolutional neural networks (CNNs). Several potential extensions and applications of the approach in other areas of machine learning and computer vision are also discussed.

In summary, the research landscape of adversarial attacks and defenses is both rich and rapidly evolving. Early works have highlighted fundamental vulnerabilities in deep learning models, while subsequent studies have proposed various defense mechanisms to counter these threats. Adversarial training and input preprocessing techniques have shown promise, but they often come with trade-offs in terms of computational cost and generalization to new attacks. Our work aims to build upon these foundational studies, exploring new strategies to enhance the robustness of image classifiers against FGSM and PGD attacks, thereby contributing to the development of more secure and reliable deep learning systems.

III. DATA AND METHODOLOGY

Our approach in this work consists of the following sequential steps: (1) Library imports and setup, (2) Data loading (3) Data preprocessing, (4) Data splitting, (5) Design of the VGG16 model, (6) Design of the convolutional autoencoder, (7) FGSM attack generation, (8) PGD attack generation, (9) Training the VGG16 model, (10) Training the autoencoder, (11) Model evaluation on accuracy, (12) Model evaluation of effectiveness, and (13) Results and analysis. The details of these steps are presented in the following.

*(1) Imports and Setup:* In this initial step, we import all necessary libraries and set up the computational environment. This includes importing libraries such as PyTorch for deep learning, NumPy for numerical computations, and Matplotlib for data visualization.

*(2) Data Loading:* We load and preprocess the dataset to be used for training and evaluation. The MNIST dataset [24] comprises 60,000 images for training and 10,000 images for testing, all depicting handwritten digits in a 28*28 grayscale format. The Fashion-MNIST dataset [25] shares a similar structure with MNIST. The Fashion-MNIST has the same structure as MNIST but contains images of clothing items such as shoes, T-shirts, and trousers.

*(3) Data Preprocessing:* All images from both datasets are resized from 28*28 pixels to 32*32 pixels to match the input size required by the VGG16 model[26]. The images are normalized to ensure the pixel values are within a specific range, typically [0, 1], and converted into tensors, which are suitable for processing by the neural network.

*(4) Data Splitting:* Both datasets are split into training and validation sets. For each dataset, 50,000 images are used for training, and 10,000 are used for validation. Data loaders are created to handle the datasets during the training, validation, and testing phases, with a batch size of 64 to ensure efficient training and evaluation.

*(5) Design of the VGG16 Model:* For implementing the VGG16 model for image classification, first a class named VGGBlock is designed. A VGGBlock object comprises two convolutional layers, succeeded by a batch normalization layer and a max pooling layer. Each convolutional layer employs a 3*3 kernel with a stride of 1 and padding of 1, ensuring the preservation of spatial dimensions post-convolution. Batch normalization, performed after each convolutional layer, aids in stabilizing and expediting training by standardizing the output of the convolutional layer. done after each convolutional layer helps stabilize and accelerate training by normalizing the output of the convolutional layer. Following the convolutional layers and the batch normalization, max pooling is applied. Max pooling reduces the spatial dimensions of the feature maps, effectively downsampling the input by taking the maximum value in each 2*2 window. This layer helps reduce the computational load and control overfitting by progressively reducing the spatial size of the data.

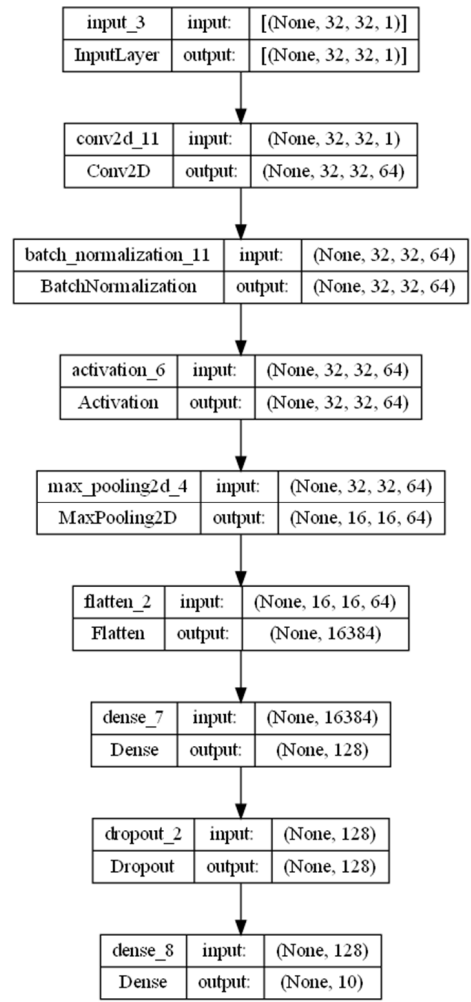

Fig. 1. The design of the VGG16 model with one illustrated convolutional layer, while the remaining subsequent convolutional layers are truncated

A VGG16 class is defined now for implementing the VGG16 neural network architecture. The network is structured with several VGGBlock instances followed by a fully connected classifier. The VGGClass consists of the following components: (1) Input Size, (2) VGGBlocks, (3) Fully Connected Classifier. The network takes an input size parameter specifying the number of channels, width, and height of the input images. The VGG16 class consists of four VGGBlock instances. Each block processes the input through its layers and passes the output to the next block. The number of output channels increases with each block, allowing the network to learn increasingly complex features. After the final VGGBlock, the output is flattened into a one-dimensional vector. This vector undergoes processing through a sequence of fully connected layers. The classifier has two main fully connected layers employing ReLU activation functions alongside dropout for regularization, followed by an output layer that produces the final class scores. During training, dropout aids in preventing overfitting by randomly deactivating a portion of the input units, and setting them to zero. Figure 1 depicts the schematic design of the VGG16 model adapted for our work. In the diagram, only one convolutional layer is shown to decrease the dimension of the figure.

*(6) Design of the ConvAutoencoder:* A convolutional autoencoder represents a neural network utilized in unsupervised learning tasks. This model is structured to acquire effective data representations, condensing the input into a latent-space representation before reconstructing it. The autoencoder crafted for our proposed scheme comprises two main components: the encoder, and the decoder.

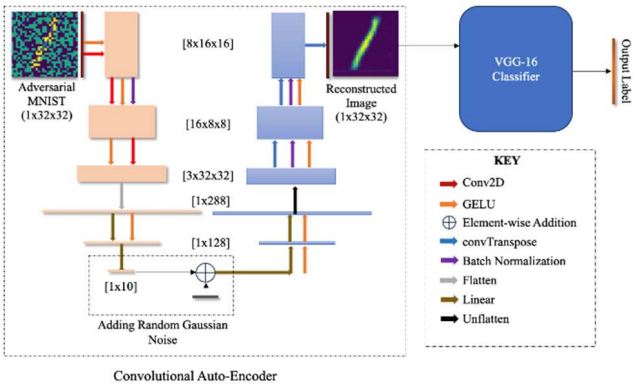

Fig. 2. The architecture of the convolutional autoencoder

Figure 2 presents the schematic design of the convolutional autoencoder. The first convolutional layer takes an input image and applies several filters to detect basic features like edges. This layer decreases the input's spatial dimensions while boosting its depth, increasing the number of feature maps. Subsequently, the activation function Gaussian Error Linear Unit (GELU) is employed to introduce non-linearity, aiding in the learning of intricate patterns [27]. GELU combines the benefits of ReLU and Gaussian noise. Unlike ReLU, which is piecewise linear and has a discontinuity at zero, GELU is a smooth function. This usually leads to a better performance as it avoids the sharp transitions that can occur with ReLU. The second convolutional layer further processes the output from the first layer, detecting more complex features. The spatial dimensions are further reduced. Batch normalization is applied to standardize the outputs of this layer, which helps in stabilizing and accelerating the training process. The third convolutional layer further extracts features and diminishes spatial dimensions, leading to a condensed depiction of the input. Following this, the output from the last convolutional layer, forming a multi-dimensional array, undergoes flattening into a one-dimensional vector. This flattened vector then traverses through a sequence of fully connected layers, further compressing the data into the latent space. The final output from these layers constitutes the latent representation.

To make the model more robust and better at generalizing, random Gaussian noise is added to the latent representation. This helps the model to learn a smoother latent space and improve its performance on unseen data.

The decoder rebuilds the input data from the latent representations through the following steps: First, the latent representation undergoes traversal through a sequence of fully connected layers, which expands it into a multi-dimensional array resembling its shape before flattening. Then, this expanded vector is reshaped into a multi-dimensional array mirroring the output shape of the last convolutional layer in the encoder. Finally, transposed convolutional layers execute operations opposite to those of the encoder's convolutional layers. They increase the spatial dimensions while decreasing the depth, gradually reconstructing the input range. Following each transposed convolutional layer, an activation function is utilized to incorporate non-linearity. Batch normalization is then employed on the output of the initial two transposed convolutional layers to stabilize the reconstruction process. Subsequently, a sigmoid activation function is applied to the ultimate output, guaranteeing that the reconstructed image possesses pixel values within the range of 0 to 1, aligning with standard image requirements.

*(7) FGSM Adversarial Attack Generation:* FGSM is employed to generate adversarial examples through subtle adjustments to the input image in a manner that amplifies the error in the model's prediction. This entails calculating the loss functions' gradient for the image and then altering the image in the direction of the gradient scaled by a small factor (epsilon).

*(8) Projected Gradient Descent (PGD) Generation:* PGD operates through iterations for generating adversarial examples [2]. It performs multiple small perturbations on the input image and ensures that the altered image stays within a designated distance (epsilon-ball) from the original image. Every iteration entails calculating the loss gradient and adjusting the image with a tiny step (*alpha*) along the gradient's path, then subsequently returning the adjusted image to the permissible epsilon-ball.

*(9) Training the VGG16 Model:* The VGG16 undergoes training on clean data extracted from the MNIST-Fashion and MNIST datasets. Throughout the training process, the model's parameters are fine-tuned employing a value of 0.001 for the learning rate, optimized using Adam, for a batch size of 64. The criterion of "early stopping" is applied to observe a loss in the validation set, ceasing the training process upon detecting indications of overfitting. This approach enhances the generalizability of the model on new data.

*(10) Training the Autoencoder:* The autoencoder is trained to reconstruct adversarial examples back to their original form. During this training, Mean Squared Error (MSE) is utilized for quantifying the mean squared disparity between the reconstructed images and their originals. The autoencoder is trained using adversarial examples generated through both FGSM and PGD methods to fortify it against various attack types.

*(11) Accuracy:* The model's accuracy is measured by the ratio of correctly classified images to the total number of

images. This metric assesses the VGG16 model's performance on both clean and adversarial datasets. This metric is used to evaluate the performance of the VGG16 model on both clean and adversarial samples.

*(12) Model Evaluation on Defense Effectiveness:* The efficacy of the autoencoder defense mechanism is assessed based on the reduction in the VGG16 model's accuracy due to the attacks. This is evaluated both before and after applying the autoencoder to the adversarial examples. The reduction in accuracy indicates the robustness of the defense mechanism.

*(13) Results and Analysis:* Finally, the model's performance results are compared on clean and adversarial examples before and after applying defense mechanisms. The model's accuracy is graphed across varying levels of perturbation to visualize the effectiveness of our proposed defense mechanism. This analysis helps quantify the improvement in the resilience of the model in the presence of the proposed defense scheme.

## IV. PERFORMANCE RESULTS

The performance of the convolutional autoencoder for defending against FGSM and PGD attacks is evaluated on the Kaggle computing platform with a P100 GPU hardware accelerator. The P100 GPU refers to the NVIDIA Tesla P100 GPU with 3584 CUDA cores16GB HBM2 RAM, 732 GB/s memory bandwidth, with 4-8 cores Intel Xeon processor.

Figure 3 shows some specimen images from the MNIST and MNIST-Fashion datasets.

Table 1 presents the classification accuracies for the VGG16 classifier during an FGSM attack with different noise levels for the datasets. Figure 4 depicts the same in graphical forms. The classifier accuracy is adversely affected by the attack.

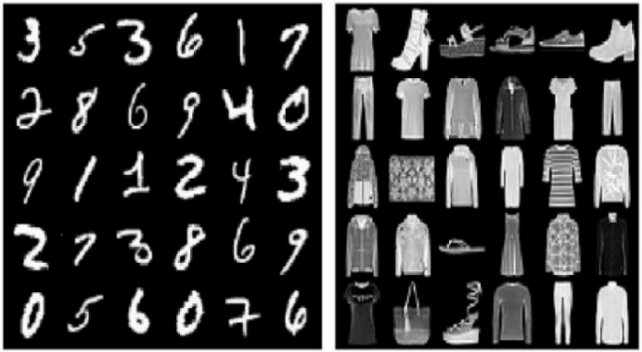

Fig. 3. Some randomly selected images from (a) MNIST dataset (left) and (b) MNIST fashion dataset ( right)

TABLE I. ACCURACY OF VGG16 MODEL UNDER FGSM ATTACK FOR DIFFERENT NOISE LEVELS ON MNIST AND MNIST FASHION TEST DATA

| Noise Level ($\varepsilon$) | Accuracy on MNIST | Accuracy on MNIST Fashion |
|---|---|---|
| 0.00 | 0.9909 | 0.9077 |
| 0.10 | 0.9077 | 0.4746 |
| 0.20 | 0.8248 | 0.3767 |
| 0.30 | 0.7278 | 0.2951 |
| 0.40 | 0.5753 | 0.2435 |
| 0.50 | 0.4139 | 0.1887 |
| 0.60 | 0.2621 | 0.1405 |
| 0.70 | 0.1611 | 0.0960 |
| 0.80 | 0.0994 | 0.0696 |
| 0.90 | 0.0714 | 0.0499 |
| 1.00 | 0.0528 | 0.0444 |

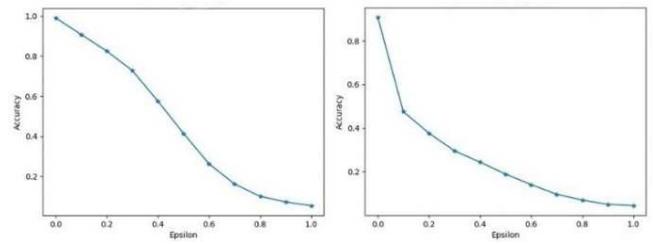

Fig. 4. Test accuracy of VGG16 model under FGSM attack for different noise levels on (a) MNIST data (on the left), (b) MNIST fashion data (on the right)

TABLE II. ACCURACY OF VGG16 MODEL UNDER PGD ATTACK FOR DIFFERENT NOISE LEVELS ON MNIST AND MNIST FASHION DATA TEST

| Noise Level ($\varepsilon$) | Accuracy on MNIST | Accuracy on MNIST Fashion |
|---|---|---|
| 0.00 | 0.9927 | 0.9603 |
| 0.05 | 0.6135 | 0.3153 |
| 0.10 | 0.1725 | 0.2947 |
| 0.15 | 0.0420 | 0.2946 |
| 0.20 | 0.0181 | 0.2943 |
| 0.25 | 0.0135 | 0.2949 |
| 0.30 | 0.0122 | 0.2956 |
| 0.35 | 0.0121 | 0.2956 |
| 0.40 | 0.0126 | 0.2948 |

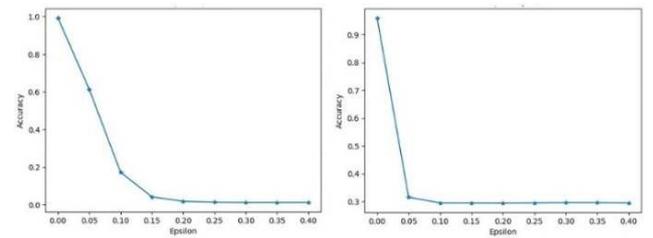

Fig. 5. Test accuracy of the VGG16 model during a PGD attack at varying noise levels on (a) MNIST data (on the left), (b) MNIST fashion data (on the right)

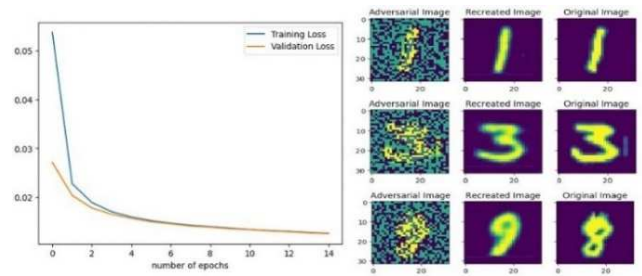

Fig. 6. (a) Training and validation loss plot of the autoencoder on adversarial samples by FGSM attack on MNIST (on the left). (b) Sample adversarial images by FGSM and their reconstructed versions by the autoencoder

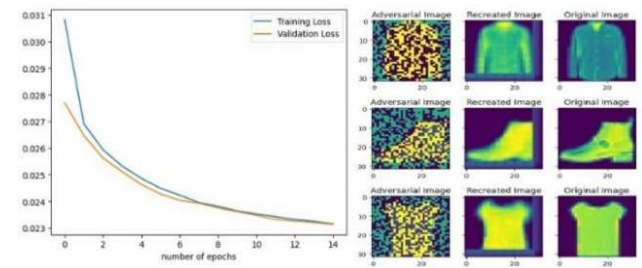

Fig. 7. (a) Training and validation loss plot of the autoencoder on adversarial samples by FGSM attack on MNIST-Fashion (on the left). (b) Sample adversarial images by FGSM and their reconstruction by the autoencoder

The impact of the PGD attack on reducing the accuracy of the VGG16 model for different noise levels is illustrated in

Figure 5 and exhibited in Table II. PGD attack has a more detrimental impact on the classifier's accuracy than its FGSM counterpart for a given value of the noise.

TABLE III. TEST ACCURACY OF VGG16 MODEL UNDER FGSM ATTACK WITH THE AUTOENCODER DEFENSE ON MNIST AND MNIST FASHION DATA

| Noise Level ($\varepsilon$) | Test Acc on MNIST | Time for defending an attack (s) | Test Acc on MNIST Fashion | Time for defending an attack (s) |
|---|---|---|---|---|
| 0.00 | 0.9569 | 0.003 | 0.7479 | 0.003 |
| 0.10 | 0.9372 | 0.0003 | 0.7063 | 0.0003 |
| 0.20 | 0.9309 | 0.0003 | 0.6582 | 0.0003 |
| 0.30 | 0.9198 | 0.0003 | 0.6652 | 0.0003 |
| 0.40 | 0.8945 | 0.0003 | 0.6652 | 0.0003 |
| 0.50 | 0.8695 | 0.0003 | 0.6558 | 0.0003 |
| 0.60 | 0.8609 | 0.0003 | 0.6703 | 0.0003 |
| 0.70 | 0.8629 | 0.0003 | 0.6866 | 0.0003 |
| 0.80 | 0.8792 | 0.0003 | 0.6988 | 0.0003 |
| 0.90 | 0.7527 | 0.0003 | 0.7175 | 0.0003 |
| 1.00 | 0.3571 | 0.0003 | 0.4605 | 0.0003 |

Figures 6 and 7 illustrate the losses during training and validation of the convolutional autoencoder designed to defend against the FGSM attack. Adversarial samples generated by the FGSM are used in the training. The figures also depict such samples and their reconstructed version by the autoencoder and their original version. Table III presents test accuracies on the MNIST and MNIST-Fashion under FGSM attack with the presence of the autoencoder defense. A comparison between the figures in Table I and Table III makes it evident that the autoencoder defense has been very effective in defending against the FGSM attack on both datasets.

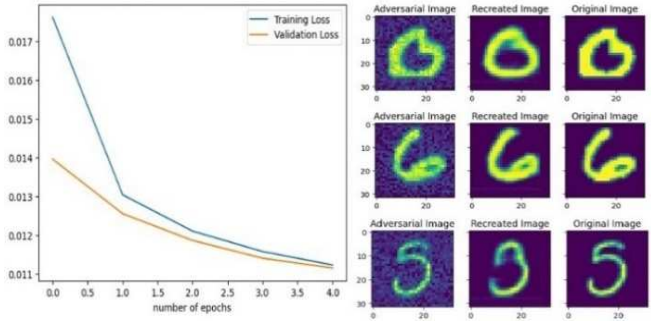

Fig. 8. (a) Training and validation loss plot of the autoencoder on adversarial samples by PGD attack on MNIST (on the left). (b) Sample adversarial images by PGD attack and their reconstructed versions by the autoencoder.

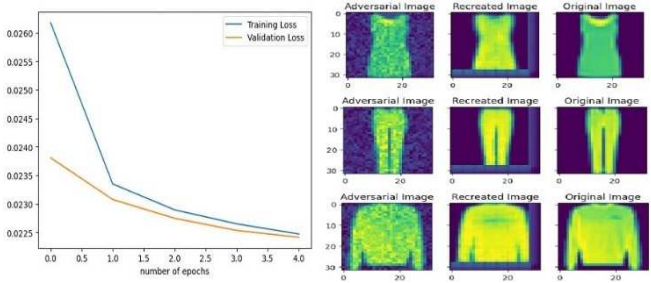

Fig. 9. (a) Training and validation loss plot of the autoencoder on adversarial samples by PGD attack on MNIST-Fashion (on the left). (b) Sample adversarial images by PGD attack and their reconstruction by the autoencoder

Figures 8 and 9 exhibit the epoch-wise loss during training and validation of the autoencoder trained to counter the PGD attack. These figures also show the adversarial samples generated by the PGD attack, their reconstructions by the autoencoder, and the original images. Table IV presents the test accuracies on the MNIST and MNIST-Fashion datasets under PGD attack in the presence of the autoencoder defense. Again, a comparison between the results in Tables II and IV reveals that the autoencoder defense significantly improves resistance to PGD attacks on both datasets.

TABLE IV. TEST ACCURACY OF VGG16 MODEL UNDER PGD ATTACK IN WITH THE AUTOENCODER DEFENSE ON MNIST AND MNIST FASHION DATA

| Noise Level ($\varepsilon$) | Test Acc on MNIST | Time for defending an attack (s) | Test Acc on MNIST Fashion | Time for defending an attack (s) |
|---|---|---|---|---|
| 0.00 | 0.9203 | 0.0012 | 0.7049 | 0.0012 |
| 0.05 | 0.9342 | 0.0012 | 0.7112 | 0.0012 |
| 0.10 | 0.9318 | 0.0012 | 0.7023 | 0.0012 |
| 0.15 | 0.9298 | 0.0012 | 0.7011 | 0.0012 |
| 0.20 | 0.9123 | 0.0012 | 0.7022 | 0.0012 |
| 0.25 | 0.9108 | 0.0012 | 0.7031 | 0.0012 |
| 0.30 | 0.9028 | 0.0012 | 0.7002 | 0.0012 |
| 0.35 | 0.9021 | 0.0012 | 0.7010 | 0.0012 |
| 0.40 | 0.9023 | 0.0012 | 0.7024 | 0.0012 |

## V. CONCLUSION

This study explored the effectiveness of a convolutional autoencoder-based defense scheme to counter FGSM and PGD attacks on MNIST-Fashion and MNIST datasets using the VGG16 classifier. The results show that the autoencoder significantly enhances the robustness of the VGG16 model by effectively reconstructing adversarially perturbed images and mitigating the impact of both FGSM and PGD attacks. Specifically, the autoencoder's preprocessing step improves classification accuracy under attack conditions, showcasing its potential as a viable defense strategy. The empirical results illustrate that the autoencoder not only defends against adversarial attacks but also maintains a high level of accuracy, even when subjected to significant perturbations. For instance, in the case of the FGSM attack, the autoencoder was able to restore much of the lost accuracy, demonstrating its ability to correct adversarial distortions in real time. Similarly, the defense mechanism proved effective against the more sophisticated PGD attack, which is known for its iterative nature and stronger adversarial perturbations.

However, while the autoencoder defense shows promise, it introduces additional computational complexity and may require fine-tuning for different datasets and attack levels. The process of training the autoencoder, especially when incorporating multiple types of adversarial attacks, can be computationally expensive and time-consuming. This suggests that there is a trade-off between the robustness provided by the defense and the computational resources required to deploy it effectively.

Moreover, the effectiveness of the defense scheme in the face of advanced attacks including the CW attack [22] or adaptive attacks, remains to be thoroughly evaluated. As adversarial attack methods continue to evolve, it is crucial to develop defense mechanisms that can anticipate and counter these more advanced techniques. Improving the scalability of autoencoder-based defenses should be a priority in future research and exploring hybrid approaches that combine multiple defense strategies for more comprehensive protection.

Looking ahead, several avenues for future work can enhance the robustness and applicability of the proposed defense mechanism. First, evaluating the autoencoder defense across a broader range of datasets, including more complex ones like CIFAR-10 and ImageNet, would be crucial in assessing its generalizability. Additionally, exploring hybrid defense strategies that combine the autoencoder with techniques such as adversarial training or defensive distillation could yield a more comprehensive and resilient approach to adversarial defense.

Addressing the computational overhead introduced by the autoencoder is another critical area, with future research focusing on optimizing the model to balance defense efficacy with practical deployment requirements. Applying this defense in real-world scenarios, such as in autonomous driving or healthcare imaging, will help determine its practical viability and uncover operational challenges that need to be addressed.

Finally, a deeper theoretical analysis of the defense's robustness and continuous monitoring of emerging adversarial threats will be necessary to keep pace with the evolving landscape of adversarial attacks and to develop adaptive defense mechanisms that can respond to new challenges.

REFERENCES

[1] I. J. Goodfellow, J. Shlens, and C. Szegedy, "Explaining and harnessing adversarial examples," *Proceedings of the International Conference on Learning Representations (ICLR'15)*, Poster Track, 2015.

[2] A. Madry, A. Makelov, L. Schmidt, D. Tsipras, and A. Vladu, "Towards deep learning models resistant to adversarial attacks," *Proceedings of the International Conference on Learning Representations (ICLR'2018)*, Vancouver, Canada, April-May 2018.

[3] N. Papernot, P. McDaniel, X. Wu, S. Jha, and A, Swami, "Distillation as a defense to adversarial perturbations against deep neural networks," *Proceedings of IEEE Symposium on Security and Privacy (SP'16)*, San Jose, CA, USA, pp. 582-597, 2016.

[4] J. Sen and S. Dasgupta, "Adversarial attacks on image classification models: FGSM and patch attacks and their impact," in J. Sen and J. Mayer (eds) *Information Security and Privacy in the Digital World: Some Selected Topics*, pp. 15-41, IntechOpen, London, UK, 2023.

[5] J. Sen, A., Sen, and A. Chatterjee, "Adversarial attacks on image classification models: Analysis and defense," *Proc. of the 10th Int. Conf, on Business Analysis and Intelligence*, Bangalore, India, 2023.

[6] N. Carlini and D. Wagner, "Adversarial examples are not easily detected: Bypassing ten detection method," *Proceedings of the 10th ACM Workshop on Artificial Intelligence and Security (AISec'17)*, pp. 3-14, November 2017.

[7] J. Li, M. Ni, Y. Dong, T. Zhu, W. Liu, "AICAttack: Adversarial image captioning attack with attention-based optimization," *arXiv: 2402.11940*, February 2024.

[8] G. Bortsova, C. Gonzalez-Gonzalo, S. C. Wetstein, F. Dubost, I. Katramadas, L. Hogeweg, B. Liefers, B. van Ginneken, J. P.W. Pluim, M. Veta, C. I. Sanchez, M. de Bruijne, "Adversarial attack vulnerability of medical image analysis systems: Unexplored factors," *Medical Image Analysis*, Vol 73, Art Id: 102141, 2021.

[9] C. Li, C. Fan, J. Zhang, C. Li, and Y. Teng, "A block gray adversarial attack method for image classification neural network," *Proc of IEEE 24th International Conference on High Performance Computing & Communications; 8th International Conference on Data Science & Systems; 20th International Conference on Smart City; 8th International Conference on Dependability in Sensor, Cloud & Big Data Systems & Applications (HPCC/DSS/SmartCity/DependSys)*, Hainan, China, pp. 1682-1689, 2022.

[10] X. Cui, A. Aparcedo, Y. K. Jang, and S-N. Lim, "On the robustness of large multimodal models against image adversarial attacks," *arXiv:2312.03777*, December 2023.

[11] F. Tramer, A. Kurakin, N. Papernot, I. Goodfellow, D. Bonesh, and P. McDaniel, "Ensemble adversarial training: Attacks and Defenses," *Proc. of International Conference on Learning Representation (ICLR'18)*, Poster Track, April-May, 2018.

[12] O. Garcia-Porras, S. Salazar-Colores, E.U. Moya-Sanchez, and A. Sanchez-Perez, "Adversarial attack versus a bio-inspired defensive method for image classification," in: M. F. Mata-Rivera et al. (eds.) *Telematics and Computing*, Communications in Computer and Information Science, Vol 1906, Springer, Cham, pp. 533-547, November 2023.

[13] A. Athalye, N. Carlini, and D. Wagner, "Obfuscated gradients give a false sense of security: Circumventing defenses to adversarial examples," *arXiv: 1802.0040*, July 2018.

[14] Y. Chen, M. Zhang, J. Li, and X. Kuang, "Adversarial attacks and defenses in image classification: A practical perspective," *Proc. of the 7th International Conference on Image, Vision and Computing (ICIVC'22)*, Xian, China, pp 424-430, 2022.

[15] M. Li and C. Cao, "Defense against adversarial attacks using image label and pixel guided sparse denoiser," *Proc of 7th International Conference on Big Data Analytics (ICBDA'22)*, Guangzhou, China, pp 253-258, 2022.

[16] N. Entezari & E. E. Papalexakis, "Tensorshield: Tensor-based defense against adversarial attacks on images," *Proc. of 2022 IEEE Military Communications Conference (MILCOM'22)*, pp 999-1004, 2022.

[17] W. Xu, D. Evans, and Y. Qi, "Feature squeezing: Detecting adversarial examples in deep neural networks," *Proc. of Network and Distributed Systems Security Symposium (NDSS)*, 2018.

[18] S. Liu, M. Shao, and X. Liu, "GAN-based classifier protection against adversarial attacks," *Journal of Intelligent & Fuzzy Systems: Applications in Engineering and Technology*, Vol 39, No 5, pp. 7085-7095, January 2020.

[19] S. Dathathri, S. Zheng, T. Yin, R. M. Murray, and Y. Yue, "Detecting adversarial examples via neural fingerprinting," *arXiv:1803.03870*, March, 2018.

[20] J. Cohen, E. Rosenfeld, and Z. Kolter, *Proc. of the 36th International Conf on Machine Learning (PMLR)*, Vol 97, pp. 1310-1320, 2019.

[21] Y.-T. Huang, W.-H. Liao, and C.-W. Huang, "Defense mechanism against adversarial attacks using density-based representation of images," *Proc. of the 25th Int Conf on Pattern Recognition (ICPR)*, pp. 3499-3504, 2021.

[22] K. Ren, T. Zheng, Z. Qin, and X. Liu, "Adversarial attacks and defenses in deep learning," *Engineering*, Vol 6, No 3, pp. 346-360, March 2020.

[23] D. Zoran, M. Chrzanowski, P.-S. Huang, S. Gowal, A. Mott, and P. Kohli, "Towards robust image classification using sequential attention models," *Proc. of IEEE/CVF Conference on Computer Vision and Pattern Recognition (CVPR'20)*, pp. 9480-9489, 2020.

[24] Y. Lecun, L. Bottou, Y. Bengio, and P. Haffner, "Gradient-based learning applied to document recognition," *Proc. of the IEEE*, Vol 86, No 11, pp. 2278-2324, November 1998.

[25] H. Xiao, K. Rasul, and R. Vollgraf, "Fashion-MNIST: A novel image dataset for benchmarking machine learning algorithms," *arXiv:1708.07747*, August 2017.

[26] K. Simonyan and A. Zisserman, "Very deep convolutional networks for large-scale image recognition," *Proc. of International Conference on Learning Representations (ICLR)*, 2015.

[27] D. Hendrycks and K. Gimpel, "Gaussian linear units (GELUs)," *arXiv:1606.08415v5*, June 2023.

[28] N. Carlini and D. Wagner, "Towards evaluating the robustness of neural networks," *Proc. of 2017 IEEE Symposium on Security and Privacy (SP)*, San Jose, USA, pp. 39-57, 2017.